\documentclass[twocolumn,prl,aps,floatfix,showpacs]{revtex4} 
\usepackage[dvips]{graphicx}
\usepackage{longtable}
\usepackage{morefloats}
\usepackage{color}
\usepackage[dvipsnames]{xcolor}
\usepackage{dsfont}
\usepackage{pdfpages}

\begin{document}

\title{Breakdown of traditional many-body theories for correlated electrons}            
\author{O.~Gunnarsson,$^1$  G.~Rohringer,$^{2}$, T.~Sch\"afer,$^3$ G.~Sangiovanni,$^4$ and A.~Toschi$^3$ } 
\affiliation{
$^1$ Max-Planck-Institut f\"ur Festk\"orperforschung, Heisenbergstra{\ss}e 1, D-70569 Stuttgart, Germany \\ 
$^2$ Russian Quantum Center, Novaya street, 100, Skolkovo, Moscow region 143025, Russia  \\
$^3$Institute of solid state physics, Vienna University of Technology, 1040 Vienna, Austria \\
$^4$ Institute of Physics and Astrophysics, University of W\"urzburg, W\"urzburg, Germany  }

\begin{abstract}
Starting from the (Hubbard) model of an atom, we demonstrate that the
uniqueness of the mapping from the interacting 
to the noninteracting Green's function, $G\to G_0$, is strongly violated, by providing
numerous explicit examples of different $G_0$ leading to the same physical $G$. 
We argue that there are indeed {\sl infinitely} many such $G_0$, with
numerous crossings with the physical solution.  We show that 
this rich functional structure is directly related to the divergence of certain classes of (irreducible vertex) 
diagrams, with important consequences for traditional many-body physics based on diagrammatic 
expansions. Physically, we ascribe the onset of these highly non-perturbative manifestations to the progressive suppression of the charge susceptibility 
induced by the formation of local magnetic moments and/or RVB states in strongly
correlated electron systems.  
\end{abstract}
\date{\today} 
\pacs{71.10.-w; 71.27.+a; 71.10.Fd}


\maketitle

{\sl Introduction.}  For more than fifty years non-relativistic quantum many-body
theory (QMBT) has been successfully
applied to describe the physics of many-electron systems in the field of condensed matter. 
Despite the intrinsic difficulty of identifying a small expansion parameter (analogously to the the fine-structure 
constant of quantum-electrodynamics), the formalism of QMBT --in its complementary
representations in terms of Feynman diagrammatics\cite{AGD,Fetter} and of the universal 
Luttinger-Ward (LW) functional\cite{Baym, Luttinger}-- is the cornerstone of the 
microscopic derivation of Landau's Fermi-liquid theory and of uncountable approximation schemes\cite{Hedin,FLEX,Bickers,fRGrev,DMFTrev}.

Yet, the actual conditions of 
applicability of the QMBT in the non-perturbative regime have been
scarcely investigated. This is surprising, because QMBT
is extensively applied to strongly
correlated electron materials, where band-theory and 
Fermi-liquid predictions fail, and some of the most exotic physics of
condensed matter systems is observed.  
Recently, however, the quest for such investigations became
particularly strong. This is because several
cutting-edge QMBT-approaches, explicitly designed for describing the crucial,
but elusive, regime of intermediate-to-strong interactions, 
have been developed, e.g., the diagrammatic
Quantum Monte Carlo (DQMC) schemes \cite{DQMC} and  numerous diagrammatic extensions\cite{DGA,DF,DB,1PI,TRILEX,QUADRILEX} of dynamical mean
field theory (DMFT) \cite{DMFTrev,DMFT}. 

Pioneering analyses of the perturbation theory breakdown 
have been reported in the
last four years\cite{Schaefer2013,Kozik,Janis2014,Stan2015,Rossi2015,Lani2012,Rossi2016,Ribic2016,landscape,Tarantino2017}. The main outcome can be summarized in two
independent observations: (i) the occurrence of {\sl infinitely} many
singularities in the Bethe-Salpeter equations and (ii) the intrinsic
multivaluedness of the LW functionals. The first problem
appears as an infinite series of unexpected {\sl divergences} in 
irreducible vertex functions \cite{landscape}, while the
second is reflected in the convergence of the perturbative series to an
unphysical solution\cite{Kozik}. 
The intrinsic
origin of these non-perturbative manifestations, their
impact on the many-electron physics, as well as on
the method development in the field, represent a challenge for the current theoretical understanding. 

In this Letter, we report a fundamental progress in the comprehension
of the perturbation theory breakdown and
of its significance. In particular, going beyond the pioneering work of \cite{Kozik}, (i) we show that there are many (probably an infinite number of)  
unphysical self-energies that become equal to the physical one at
specific values of the interaction. 
This puts us into the position 
to (ii) demonstrate the actual
correspondence between the vertex divergences of the Feynman
diagrammatics  and the occurrence of bifurcations
of the LW functional. Finally, (iii) we generalize these results from the Hubbard atom to 
generic systems with strong correlations. Regarding the nature of the singularities, we  show that vertex divergences of different kinds
are reflected in different natures of solution crossings at the bifurcation. 

The emerging scenario, which {\sl mathematically} depicts  an unexpectedly complex
structure of the many-body formalism, will be {\sl physically} related to the
progressive suppression of charge fluctuations, a generic property of strongly-correlated systems with a local interaction. 
The improved understanding
of the QMBT beyond the perturbative regime  serves as a crucial guidance  
for future method development for non-relativistic many-electron
systems.

\begin{figure}
{{\resizebox{6.0cm}{!}{\includegraphics {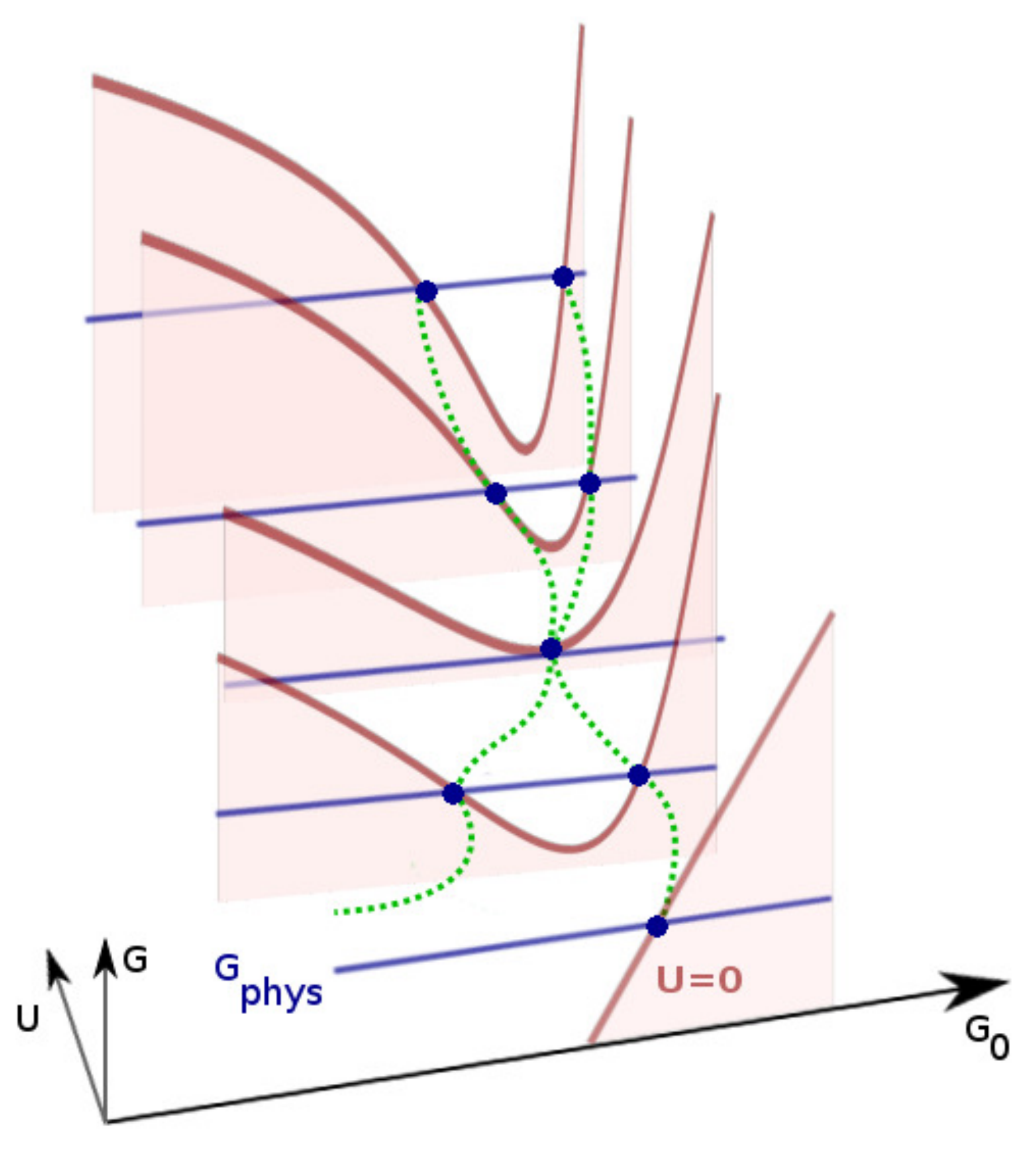}}}}
\caption{Sketch of the functional $G[G_0,U]$, where $G_0$ and $G$ are assumed to be just numbers: The red curves correspond to cuts for different values of $U$, the horizontal blue lines show the corresponding values of $G_{\text{phys}}$, while the blue dots represent the $G_{0}$ which produce $G_{\text{phys}}$.
}\label{fig:1}
\end{figure}

{\sl Multivaluedness of the Luttinger-Ward functional.}  The
Luttinger-Ward (LW) functional $\Phi[G]$ plays 
a crucial role in traditional many-body 
physics \cite{Luttinger}. It is a universal functional of the full single-particle Green's 
function $G$, which only depends on the electron-electron interaction but not on the external 
potential. From $\Phi[G]$ the free energy can be determined. One can also obtain the electron 
self-energy $\Sigma[G]\sim  \delta \Phi[G]/ \delta G$, entering the 
Dyson equation 
\begin{equation}\label{eq:0}
G_0^{-1}-G^{-1}=\Sigma,
\end{equation}
where $G_0$ is the noninteracting Green's function determined by the external potential.
From $\Sigma[G]$ one can compute all irreducible vertices $\Gamma$ entering the Bethe-Salpeter 
equations \cite{Bickers} for response functions. For instance, the charge 
susceptibility is determined by the vertex \cite{Baym, Bickers, landscape, Kozik} 
\begin{equation}\label{eq:1}
\Gamma_c={\delta \Sigma[G]\over \delta G}.
\end{equation} 
Approximations built within this approach are guaranteed to
be conserving \cite{Baym}, therefore it is exploited for numerous formal derivations \cite{formal,DMFTrev,Maier, diagram}.
Moreover, the full two-particle nature of the vertices $\Gamma$
represents an ideal building block for approximations designed to
preserve the Pauli principle properties, and related sum rules. 
In this respect, it  is believed that the parquet
equations \cite{Bickers} are one of the most
fundamental ways  of performing diagrammatic  summations.

\begin{figure*}
{\rotatebox{-90}{\resizebox{6.0cm}{!}{\includegraphics {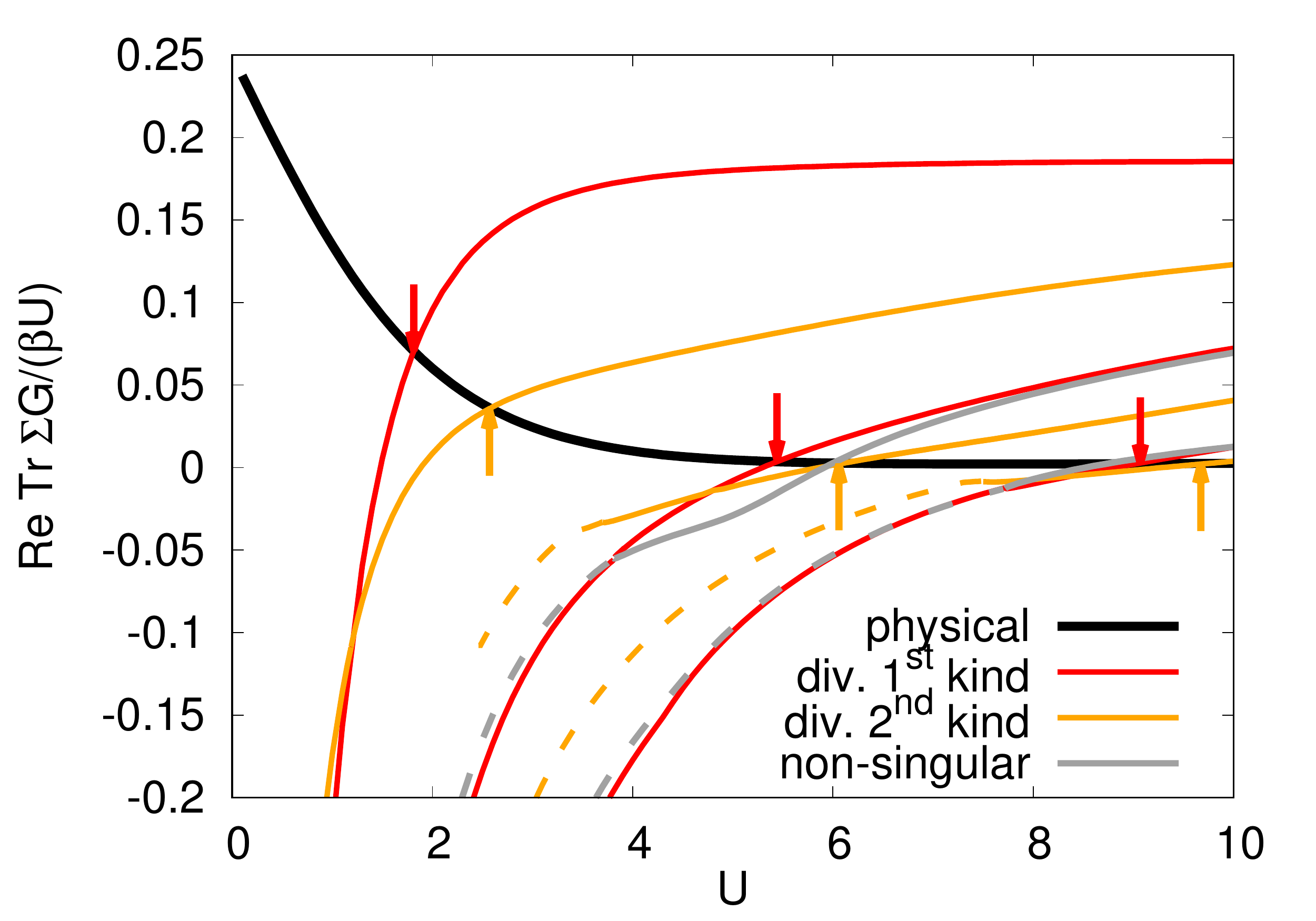}}}}
\rotatebox{-90}{\resizebox{6.cm}{!}{\includegraphics {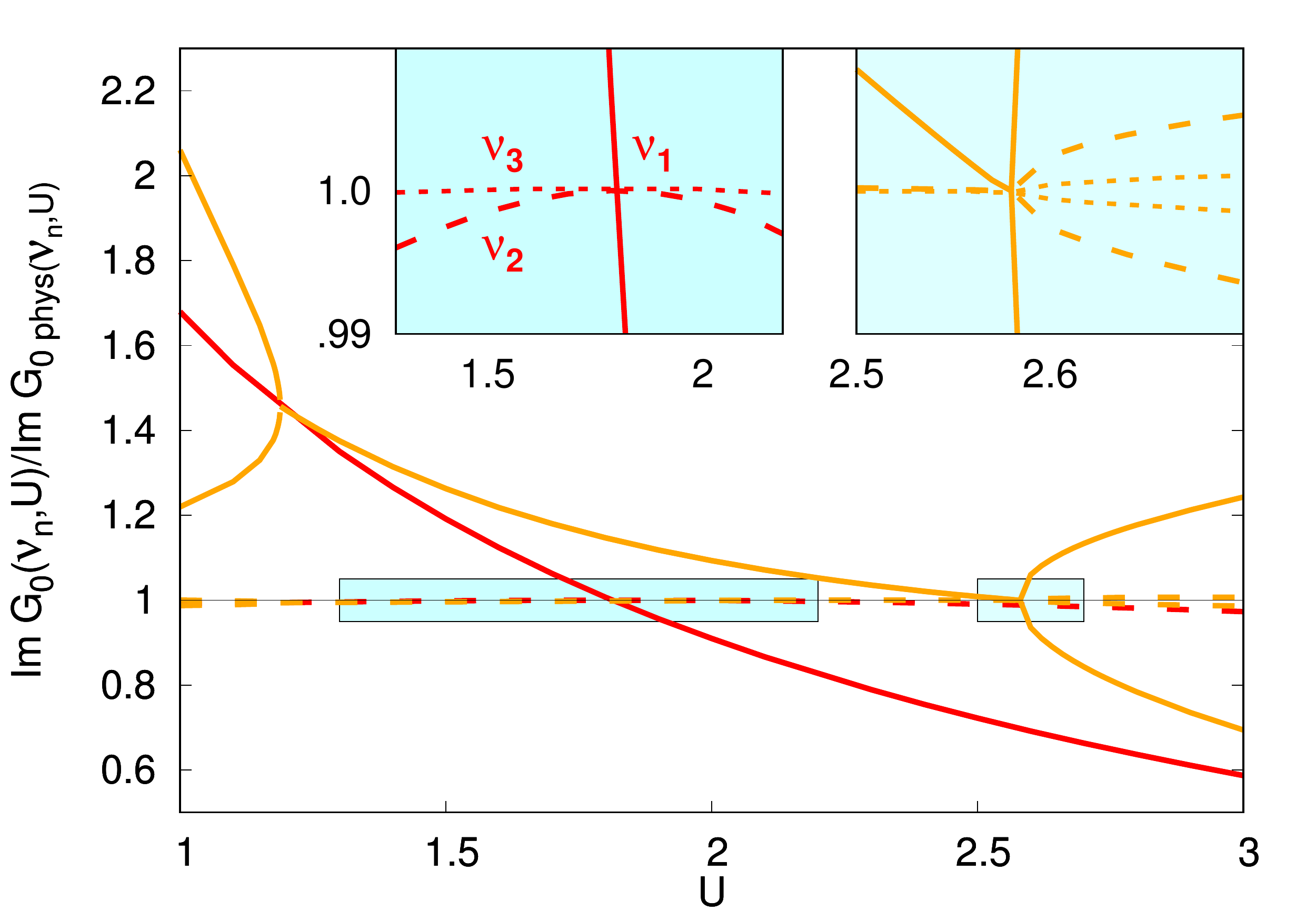}}}
\caption{Left: Tr $\Sigma G/(\beta U)$ as a function of $U$ corresponding to different 
$G_0$ for the Hubbard atom for $\beta=2$. Red/orange arrows 
mark the divergences of $\Gamma_c$ according to Ref.~\cite{landscape}. 
The physical curve corresponding to $G_{0}^{\rm phys}$ is shown in
black.  Beyond the red/orange ones, one finds further crossings (gray lines) for which, however, $G_0(\nu) \neq
G_0^{\rm phys}(\nu)$ and, hence no divergence of $\Gamma_c$ is found (see Sec.~IC of the Supplemental Material \cite{suppl}). 
Dashed curves indicate that
Tr $\Sigma G/(\beta U)$ also has an imaginary contribution, which is
not shown. Right: imaginary part of $G_0$ (normalized to Im $G_{0}^{\rm phys}$) for the first red
  and orange  curves. Results for the absolutely lowest (bold lines), second lowest (dashed lines) and (in insets)
third lowest (dotted lines) $\nu_n$ are shown. 
\label{fig:2} }
\end{figure*}

In order for QMBT methods to be meaningful, an important property of the functional $G[G_{0}]$ needs to be fulfilled: The                
introduction of $\Sigma$  in QMBT implicitly assumes that there is a 
unique mapping between $G$ and $G_0$, $G\to G_0$. 
Otherwise, several branches of $\Sigma$ would exist, corresponding to
different $G_0$, posing the general problem of an intrinsic
multivaluedness of any QMBT-based scheme. This is not just a formal 
issue. If two such branches 
cross, $\Gamma$ in Eq.~(\ref{eq:1}) might become ill-defined and
diverge. This would challenge important aspects of the
traditional many-body theory, such as, e.g., the  definition of physically meaningful
parquet summations \cite{Gunnarsson2016}.

Fig.~\ref{fig:1} schematically illustrates such a scenario.
The general functional relation between $G$ and
$G_0$ is depicted by several red curves for different values of the electronic interaction $U$ \cite{footnote_OPM}.
$G$ and $G_0$ are here treated as numbers rather than functions (of
frequency/momentum/spin, etc.). For $U=0$,  
 $G[G_0]= G_0$, and for any physical $G$
(horizontal blue line) the corresponding $G_0$ is {\sl  univocally}
determined. When $U>0$,  however,                         
$G[G_0]$ becomes ``wavier'', displaying several
maxima/minima in the functional space. This way, the intersection with
$G_{\rm phys}$ would correspond to several $G_0$ (blue dots), 
of which only one describes the physical system ($G_0^{\rm phys}$). 
Even if unphysical $G_{0}$'s exist, many standard numerical algorithms are able to converge to the solution that is adiabatically connected with the $U=0$ one. 
This can however turn into an actual problem, {\sl if} for some values of $U$ the
intersection with $G_{\rm phys}$ occurs at one extreme of
$G[G_0]$. 
This would correspond to
the intersection of two different solutions of $G_0$ (and thus of
$\Sigma$, see green dashed lines in Fig.~\ref{fig:1}). 
At this point we would expect $\delta G/\delta G_0=0$. Combining this with the Dyson
equation and the definition [Eq.~(\ref{eq:1})] of $\Gamma_c$, one would conclude that 
$\Gamma_c$ diverges. 

To go beyond the sketch of Fig.~\ref{fig:1},  
we present calculations for the Hubbard atom \cite{Hubbard} and show that different $G_0$ indeed 
do cross for certain values of $U$. In Sec.~IA of the Supplemental Material \cite{suppl} we then show that
such a crossing indeed does lead to divergences of $\Gamma_c$.

{\sl Method.}                         
We have developed a method for finding different $G_0$'s which lead to the physical $G$ for 
the Hubbard atom. We use the Hirsch-Fye algorithm \cite{HirschFye} to 
obtain $G$ from a guess for $G_0$. This method involves a summation over auxiliary
spins, which is  usually done stochastically. Here we perform a
complete summation using the Gray code \cite{Gray}, thereby avoiding stochastic 
errors. We guess a $G_0$, and then use a Metropolis method to search for improved 
guesses for $G_0$. When a promising guess has been found, the Hirsch-Fye equations 
are repeatedly linearized and solved, until a $G_0$ has been found which
accurately reproduces the physical $G$ (see Sec.~II of \cite{suppl}). It is crucial that there are 
no stochastic errors in this approach. The method makes it possible to 
determine if two $G_0$ really become equal for some $U$ and to determine
how they approach each other as $U$ is varied. 

{\sl Results for the Hubbard atom.}
We start to present our results by showing in Fig.~\ref{fig:2} (left panel) Tr $\Sigma G_{\rm
  phys}/(\beta U)$ as a function of $U$ corresponding to 
the different $G_0$ and, hence $\Sigma$, via Eq.~(\ref{eq:0}).
For $G_0 = G_{0}^{\rm phys}$, this quantity yields the double 
occupancy. The black curve displays the values obtained
with $G_0= G_0^{\rm phys}$, while the colored (red, orange) curves are the results
 for the other (unphysical) $G_0$,
 collapsing to $G_0^{\rm phys}$ in the several crossing points
 shown in the Figure. 
The latter ones {\sl do} coincide -within our numerical accuracy-  with the
locations (marked by vertical arrows) of the first six divergences of $\Gamma_c$ in the Hubbard atom \cite{landscape}.
The red $G_0$'s are associated with a milder
violation of physical constraints than the orange ones (e.g., the former
can acquire a non-zero real part, but the latter
can even violate the generic condition: $G_0(\nu)=G_0(-\nu)^{*}$ \cite{suppl}).
The left red curve was found by Kozik {\it et al.} \cite{Kozik}, although, there, it could 
not be converged around the crossing with $G_0^{\rm phys}$.

There are important connections between the frequency dependence of
the divergences of $\Gamma_c$, coded 
 by red/orange colors \cite{landscape}, and the detailed behavior of $G_0$ at a crossing.   
The divergences of $\Gamma_c$ can be divided into 
 two classes \cite{Schaefer2013,landscape}. 
To this end, we consider the generalized
 charge susceptibility $\chi_c^{\nu\nu'(\omega=0)}$
 \cite{Rohringer2012}, which
 depends  on two fermionic Matsubara frequencies, $\nu$ and $\nu'$, and a bosonic frequency $\omega=0$. 
$\Gamma_c$ is then given by 
\begin{equation}\label{eq:3}
\Gamma_c=\beta^2[ \chi_c^{-1}-\chi_{0}^{-1}],
\end{equation}
where $\chi_{0}$ is the noninteracting generalized susceptibility $\chi_c$, and $\chi_c$ and $\chi_{c,0}$ are treated 
as matrices of $\nu$ and $\nu'$. These matrices can be diagonalized 
\begin{equation}\label{eq:4}
\chi_c^{\nu\nu'(\omega=0)}=\sum_l V_l(\nu)^{*}\varepsilon_l V_l(\nu'),
\end{equation}
with $V_l(\nu)$ and $\varepsilon_l$ being  the corresponding
eigenvectors/-values. While the divergences of $\Gamma_c$ always
correspond to the vanishing of one $\varepsilon_l$, they differ in
the frequency-structure of $V_l(\nu)$: For the
divergences marked by the red arrows, $V_l(\nu)$ has only two non-zero
elements (at $\nu = \pm \nu_n=\pm(2n-1)\pi/\beta$, with $n=1,2,3, ...$) reflecting a {\sl
  localized} divergence at $\nu= \pm \nu_n$, while for the
orange arrows,  $V_l(\nu) \neq 0 \, \forall \nu$, reflecting a {\sl
  global} divergence of $\Gamma_c$ \cite{landscape}.

An analogous classification is also applicable to the different
$G_0$ resolved in frequency space: 
In Fig.~\ref{fig:2}b  we plot the ratio ${\rm Im} \ G_0(\nu)/{\rm Im} \ G_0^{\rm phys}(\nu)$ corresponding to the first
two crossings and for the three lowest $\nu$. As $G_0^{\rm phys}$ is purely imaginary, the condition 
  $G_0(\nu)=G_0^{\rm phys}(\nu)$ is reflected in their ratio being $1$  
  and ${\rm Re} \ G_0/{\rm Im} \ G_0^{\rm phys}=0$ (shown in Supplemental Material \cite{suppl}).
Fig.~\ref{fig:2}b demonstrates that the red/orange crossings observed in Tr $\Sigma
  G_{\rm phys}$  indeed corresponds to an actual identity of
  $G_0(\nu)=G_0^{\rm phys}(\nu),\, \forall \nu$ both at $U=1.81$  (red) and $2.58$
  (orange). Yet, the corresponding zooms in the insets show a
  {\sl qualitative} difference between the two cases. For the red case, 
  the crossing of $G_0$ with $G_0^{\rm phys} $ is {\sl linear}  in
  $U$  only for $\nu= \pi/\beta$ (solid line), while it is $O(U^2)$ for all other
  $\nu_n$ (dashed line). In the orange case, the crossings display the {\sl same} behavior for all the frequencies (see insets of Fig.~\ref{fig:2} and 
  the discussion in Sec.~IA of \cite{suppl}). This leads to a divergence of 
$\Gamma_c$ for $\nu, \nu'=\pm \pi/\beta$ at the red crossing and for all frequencies 
at the orange crossing.
Similar results are found for the second and third red/orange crossings,
but for the former the linear crossing happens for
$\nu_2=3\pi/\beta$ (second) and $\nu_3=5\pi/\beta$ (third).  
This is consistent with the result in 
Ref.~\cite{landscape} that the corresponding divergences of $\Gamma_c$ happens at
these specific $\nu_n$. 

The one-to-one correspondence of red/orange crossings with the
 local/global divergences of $\Gamma_c$ illustrates  how the heuristic
scenario of Fig.~\ref{fig:1} is actually realized for the Hubbard
atom. The result also indicates the existence of an {\sl infinite} number of unphysical  $G_0$, since              
{\sl infinitely} many red and orange
divergences were found for the Hubbard atom\cite{landscape}.
Furthermore, there are indications that the infinity of the total number of $G_0$ 
might be of an higher cardinality than that of the  vertex
divergences, as we discuss  in Sec.~III of the Supplemental Material \cite{suppl}.

{\sl Generic strongly correlated systems.} --  To make closer contact to strongly correlated
physical systems, we consider the Hubbard model in DMFT \cite{DMFT,DMFTrev}, 
where the Hubbard atom is embedded in a self-consistent,
noninteracting host.
 The LW  functional is unchanged, since it only depends on the interacting part of the 
Hamiltonian. The external-potential part, however, changes, and therefore both 
$G_{\rm phys}$ and $G_{0}^{\rm phys}$ are different.
We can exploit the relation of the crossings with the divergences
of $\Gamma_c$ and the zero eigenvalues in Eq.~(\ref{eq:4}), and  gain further insight 
by analyzing the physical local charge susceptibility. In DMFT, this is given by \cite{Gunnarsson2016}
\begin{equation}\label{eq:6}
\chi_{\rm ch}={1\over \beta^2}\sum_{\nu\nu'} \chi_c^{\nu,\nu'(\omega=0)}
={1\over \beta^2}\sum_l\varepsilon_l |\sum_{\nu} V_l(\nu)|^2.
\end{equation}

\begin{figure}
{\rotatebox{-90}{\resizebox{6.cm}{!}{\includegraphics {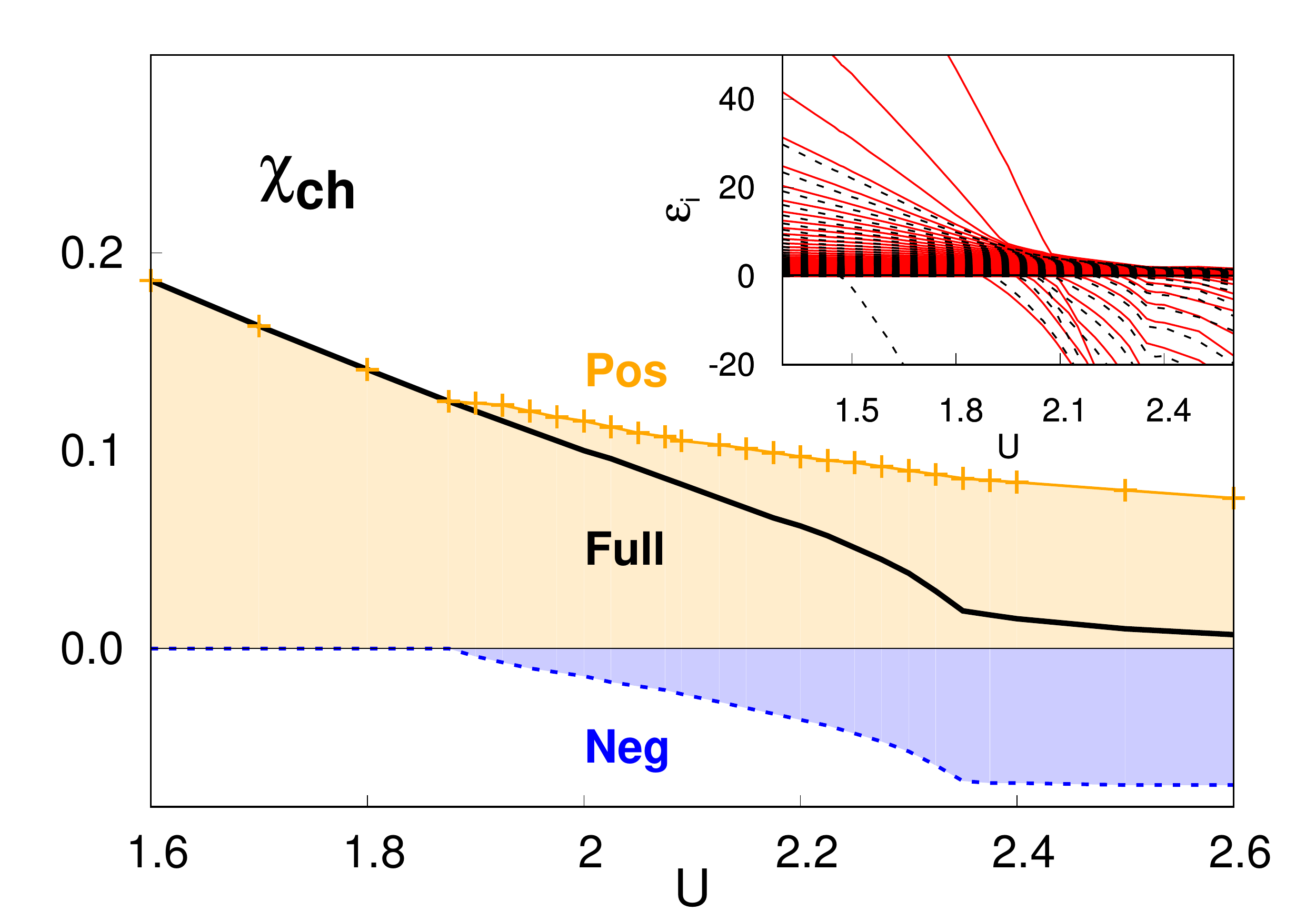}}}}
\caption{DMFT calculation of the local charge susceptibility
  ($\chi_{\rm ch}$) [Eq.~(\ref{eq:6})] (black line) as a function of $U$ for a
  two-dimensional Hubbard model at half-filling, $\beta=40$ and $4t=1$. 
Both $\chi_{\rm ch}$ and its contributions from positive (orange)
and negative (blue) eigenvalues are shown. The inset shows the $U$-dependence of the eigenvalues
$\varepsilon_i$, using solid lines for $\varepsilon_i$ with 
$\sum_{\nu} V_l(\nu)\ne 0$ and dashed lines for the rest. 
}\label{fig:3}
\end{figure}

The corresponding DMFT results are reported in
Fig.~\ref{fig:3}. By increasing $U$, the electrons gradually  
localize, building up local magnetic moments with longer lifetimes. These,
in turn, freeze the local charge  dynamics, with $\chi_{\rm ch}$ becoming very small
especially in the proximity/after the  Mott metal-insulator
transition (between $U=2.3$ and $U=2.4$ for $\beta=40$).
While the physics of this generic trend is known, the projection
of $\chi_{\rm ch}$ in its eigenvalue-basis yields highly
non-trivial information. We 
analyze  $\chi_{\rm ch}$ in terms of the contributions from  
positive and negative $\varepsilon_l$. 
For small/moderate $U$ all $\varepsilon_l$ are positive. As $U$ increases, one $\varepsilon_l$
after the other goes through zero (see inset of Fig.~\ref{fig:3}).
Each time $\Gamma_c$ diverges [see Eqs.~(\ref{eq:3},
\ref{eq:4})],  a new $G_0$ becomes identical to $G_{0}^{\rm phys}$, and
the negative component of $\chi_{\rm ch}$ becomes more important.
Such a negative
component of $\chi_{\rm ch}$  plays a crucial role in realizing the correct strong coupling
physics. Its neglection would lead to a $\chi_{\rm ch}$
approximately saturating at some sizable value in the Mott phase,
instead of being strongly suppressed.
We also note, that a small value of $\chi_{\rm ch}$ in itself is not
sufficient for this scenario to be realized (e.g., in a dilute system 
$\chi_{\rm ch}$ can be small, but all $\varepsilon_l>0$). Here, the
crucial factor is the mechanism responsible for the reduction of
$\chi_{\rm ch}$: the gradual local moment formation which manifests itself
in a progressively larger contribution of the negative $\varepsilon_l$.
This is, thus, the underlying physics responsible for the
occurrence of the (infinitely many) unphysical $G_0$ crossing
$G_{0}^{\rm phys}$, and  the related divergences of $\Gamma_c$.
This also applies to the corresponding breakdown of  perturbative
expansions, such as parquet-based approximations. 
A certain class of diagrams can give a positive infinite contribution, 
which is canceled by another class of diagrams \cite{Gunnarsson2016}. 
Then the diagrammatic expansion is not absolutely convergent, as was 
also found in Ref.~\cite{Kozik}. This makes conventional diagrammatic expansions highly  
questionable for intermediate-to-strong correlations. 
While it is not surprising that  perturbative approaches
might break down at the Mott transition,  it is interesting that this
happens well before the Mott transition occurs, where 
Fermi-liquid physical properties still control the low-energy
physics.

It is important to stress that the non-perturbative
manifestations discussed in this Letter are affecting not {\sl only} models
dominated by purely local physics (such as the Hubbard atom or its DMFT
version). In fact, divergences of $\Gamma_c$ have also been 
found \cite{Gunnarsson2016} studying the $2d$ Hubbard model, by means of the dynamical 
cluster approximation (DCA) \cite{Maier}. In this case,  the
underlying physics behind the change of sign of the $\varepsilon_l$ could be related to the formation of an RVB
state \cite{Liang88}, also responsible \cite{Jaime} for the opening of a spectral pseudogap \cite{Timusk}. 
Increasing the size of the DCA cluster might even push the occurrence of the
first  $\varepsilon_l=0$ and the pseudogap towards
lower $U$, due to strong antiferromagnetic fluctuation extended on larger length scales \cite{Schaefer2015, Gunnarsson2016, FluctDiag}.

{\sl Conclusions.} -- We have reported important progress towards the 
understanding of the mathematical structures of quantum many body theories
in the non-perturbative regime, and of the physics behind them. The
structure of the LW functionals is even richer than the pioneering
work by Kozik {\sl et al.} suggested \cite{Kozik}: We find a very large,
probably {\sl infinite}, number of noninteracting $G_0$ leading to
the same dressed $G$. This can be regarded as a formal problem,
as long as the unphysical and physical $G_0$ do not intersect, as it is
the case in the perturbative regime. However, in the nonperturbative
regime  we find many crossings. These crossings reflect the
analytical structure of the LW functional for physical systems. We
show that they lead to  divergences of irreducible vertex functions.
This challenges current quantum many-body algorithms in several
respects, e.g., causing non-invertibility of the Bethe-Salpeter
equation and breakdown of the parquet resummations. These problems, which occur when the correlation is 
still substantially weaker than in a Mott insulator, are, nonetheless, originated in underlying strong-coupling physical mechanisms,
e.g. the formation of local moments and RVB states. This is reflected in the progressive  
suppression of the charge susceptibility in correlated systems. Further investigations of the theoretical
foundations  beyond the perturbative regime should play a central role
for future method developments in condensed matter physics.

{\sl Acknowledgments.} --  We thank S. Ciuchi,  P. Thunstr\"om, 
M. Capone, S. Andergassen and K. Held for insightful discussions and P.~Chalupa also for carefully reading the manuscript. 
The authors would like to thank all attendees to the
Workshop ``Multiple Solutions in Many-Body Theories''
held in Paris the 13$^{\text{th}}$ and 14$^{\text{th}}$ of June 2016 for interesting
discussions.
We acknowledge support from FWF through the projects I-2794-N35 (TS, AT) and
from the research unit FOR 1346 of the DFG (GS).

\pagestyle{empty}

\clearpage

\noindent
\hskip -12mm
\includegraphics[width=1.10\textwidth]{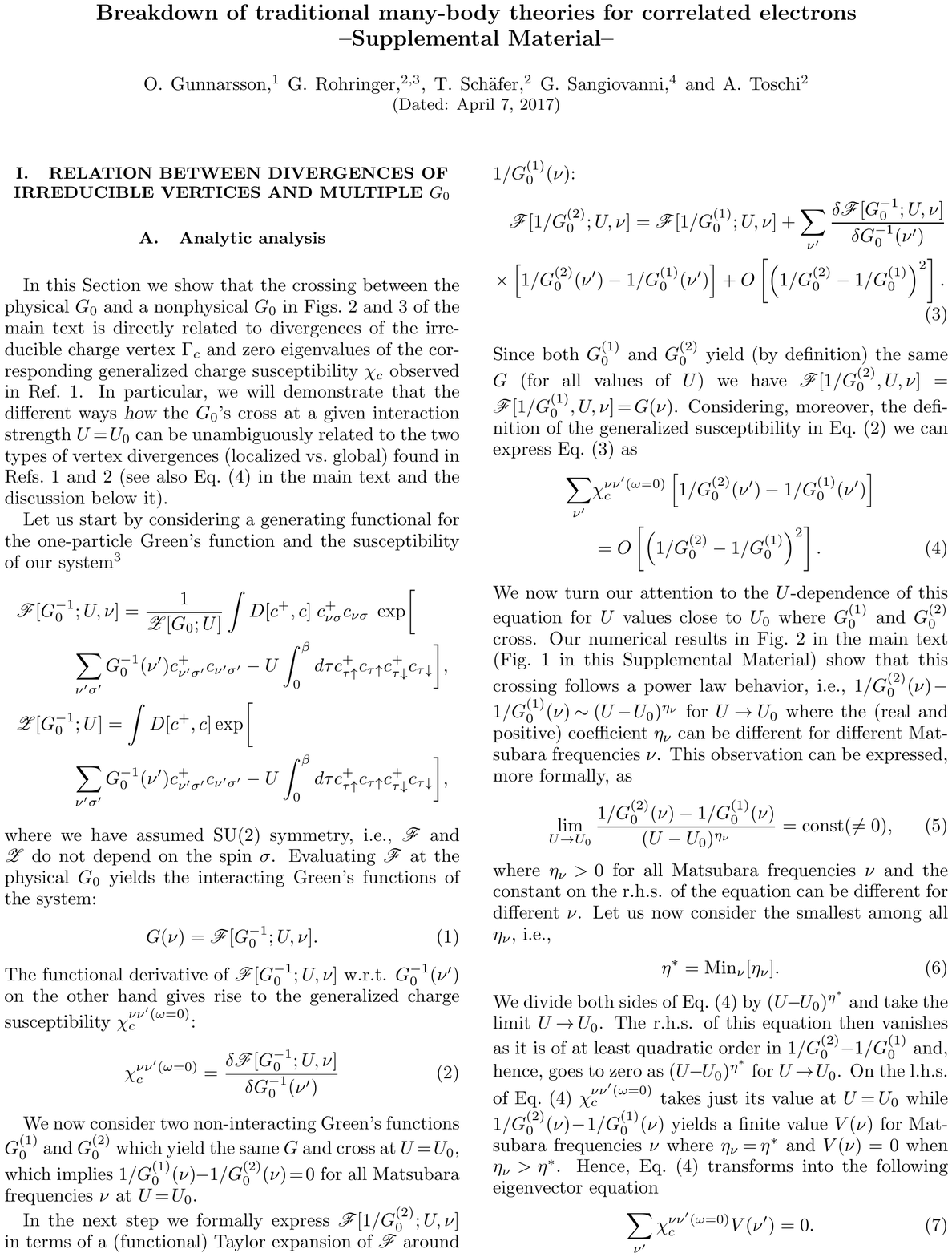}

\clearpage

\hskip -12mm
\includegraphics[width=1.10\textwidth]{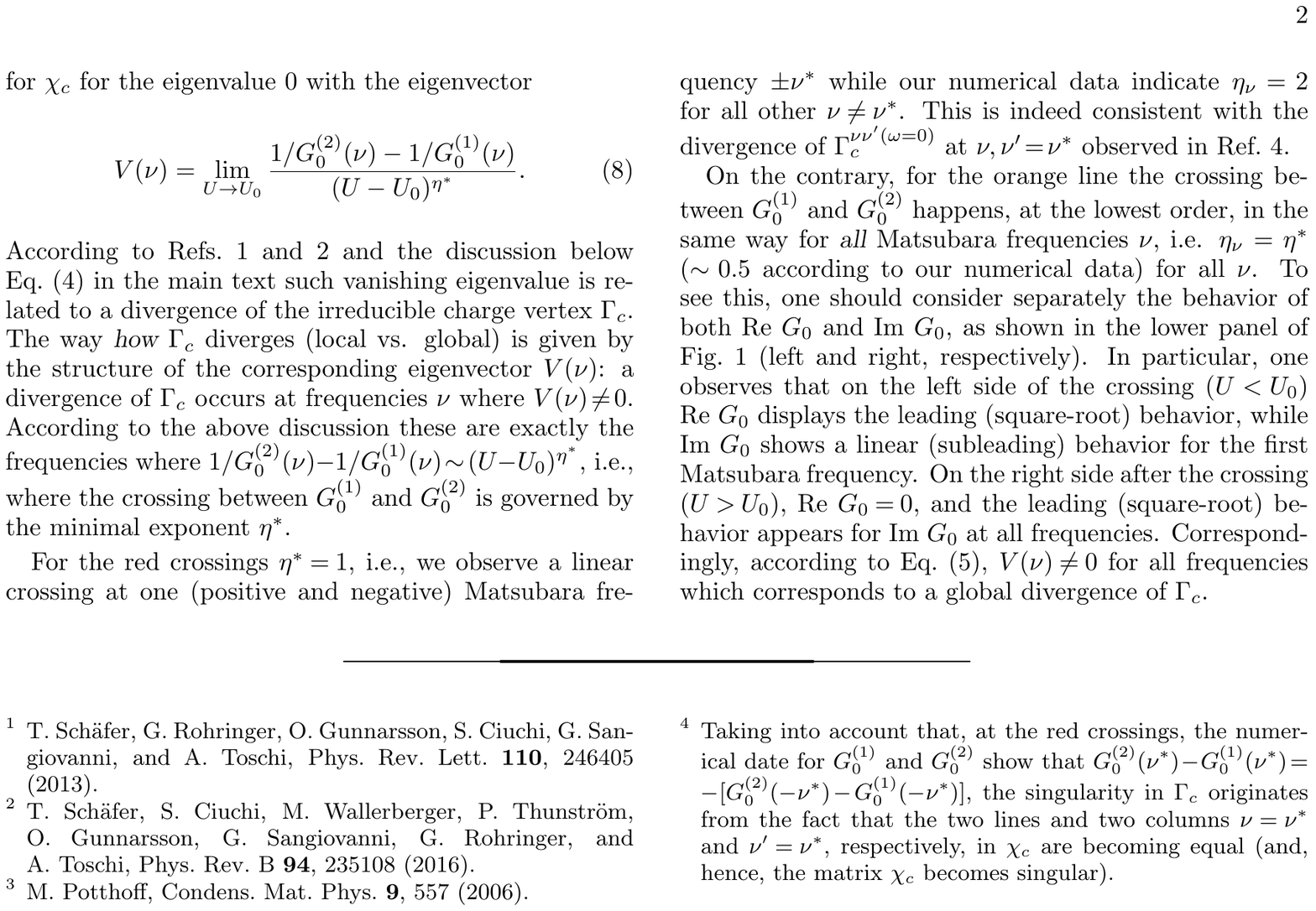}

\end{document}